\documentclass[prd,twocolumn,superscriptaddress,nofootinbib]{revtex4-1}
\usepackage{amsmath,amssymb,mathtools}
\usepackage{graphicx}
\usepackage{xcolor}
\usepackage{multirow}
\usepackage{cancel}
\usepackage{color}
\usepackage{ulem}
\usepackage{enumitem}

\usepackage[colorlinks,citecolor=blue]{hyperref}

\newcommand{\beq}{\begin{equation}}
\newcommand{\eeq}{\end{equation}}
\newcommand{\bea}{\begin{eqnarray}}
\newcommand{\eea}{\end{eqnarray}}
\newcommand{\besp}{\begin{equation}\begin{split}}
\newcommand{\eesp}{\end{split}\end{equation}}
\newcommand{\met}{\not{\!\!{\rm E}}_{T}}
\newcommand{\nn}{\nonumber}

\newcommand{\tabincell}[2]{\begin{tabular}{@{}#1@{}}#2\end{tabular}}

\begin{document}

\title{Searching for Weak Singlet Charged Scalar at the Large Hadron Collider}

\author{Qing-Hong Cao}
\email{qinghongcao@pku.edu.cn}
\affiliation{Department of Physics and State Key Laboratory of Nuclear Physics and Technology, Peking University, Beijing 100871, China}
\affiliation{Collaborative Innovation Center of Quantum Matter, Beijing 100871, China}
\affiliation{Center for High Energy Physics, Peking University, Beijing 100871, China}

\author{Gang Li}
\email{gangli@phys.ntu.edu.tw}
\affiliation{Department of Physics, National Taiwan University, Taipei 10617, Taiwan}
\affiliation{Department of Physics and State Key Laboratory of Nuclear Physics and Technology, Peking University, Beijing 100871, China}

\author{Ke-Pan Xie}
\email{kpxie@pku.edu.cn}
\affiliation{Department of Physics and State Key Laboratory of Nuclear Physics and Technology, Peking University, Beijing 100871, China}

\author{Jue Zhang}
\email{juezhang87@pku.edu.cn}
\affiliation{Center for High Energy Physics, Peking University, Beijing 100871, China}

\begin{abstract}

Weak singlet charged scalar exists in many new physics models beyond the Standard Model. In this work we show that a light singlet charged scalar with mass above 65~GeV is still allowed by the LEP and LHC data.
The interactions of the singlet charged scalar with the Standard Model particles are described by operators up to dimension-5. Dominant decay modes of the singlet charged scalar are obtained, and a subtlety involving field redefinition and gauge fixing due to a dimension-5 operator is also clarified. We demonstrate that it is promising to observe the singlet charged scalar at the LHC. 
\end{abstract}

\maketitle

\section{Introduction} 

Charged scalar is an undoubted signal of new physics (NP) beyond the Standard Model (SM). Many searches for charged scalars have been carried out at colliders and most of them focus on the charged scalars arising from either weak doublets or triplets. Doublet charged scalars have been widely discussed, and a prominent example would be the charged Higgs in the Two Higgs Doublet Model (THDM)~\cite{Lee:1973iz,Branco:2011iw}. A combined analysis of all Large Electron Positron Collider (LEP) results shows that such a charged Higgs is excluded up to a mass of around $80~\mathrm{GeV}$~\cite{Abbiendi:2013hk}. The discussion on the weak triplet case, such as doubly charged scalars in the Georgi-Machacek model~\cite{Georgi:1985nv} and Type-II seesaw models~\cite{Magg:1980ut,Cheng:1980qt,Schechter:1980gr,Lazarides:1980nt,Mohapatra:1980yp}, can be found, e.g., in Refs.~\cite{kang:2014jia,Perez:2008ha,Chao:2008mq,Han:2015hba}. Unfortunately, a dedicated study on the collider phenomenology of the weak-singlet singly-charged scalar $S$ is still missing, especially it is unclear whether the above lower bound for the doublet charged Higgs also applies to the singlet case.

\begin{figure}
\centering
\includegraphics[scale=0.33]{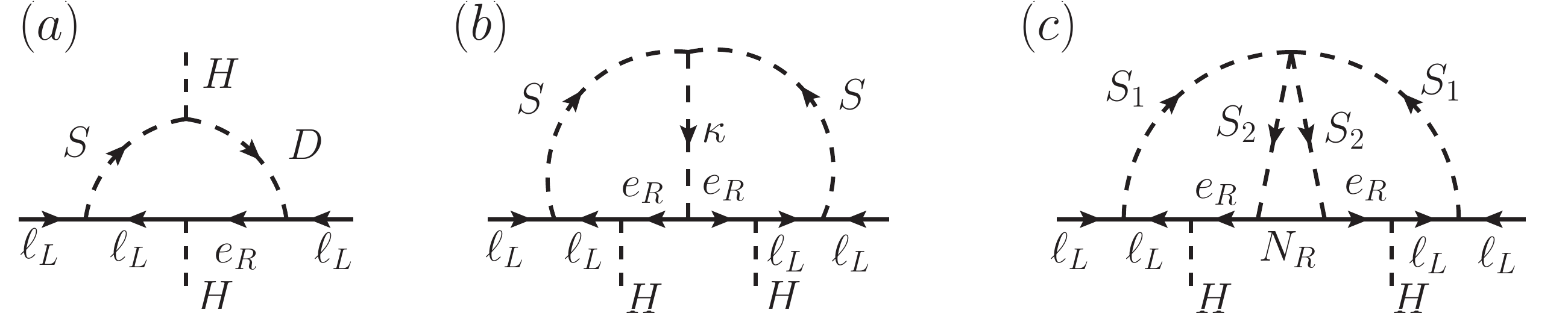}
\caption{Examples of radiative neutrino mass models that contain the singlet charged scalar $S\sim(\mathbf{1},\mathbf{1})_{-1}$. The gauge quantum numbers of the SM fields are: $H \sim (\mathbf{1},\mathbf{2})_{1/2}$, $\ell_{\rm L}^{} \sim (\mathbf{1},\mathbf{2})_{-1/2}$ and $e_{\rm R}^{} \sim (\mathbf{1},\mathbf{1})_{-1}$. (a) The one-loop model from Ref.~\cite{Zee:1980ai}, where $D \sim (\mathbf{1},\mathbf{2})_{1/2}$ is another scalar doublet. (b) The two-loop model from Ref.~\cite{Aoki:2010ib}, where a singlet doubly charged field $\kappa\sim(\mathbf{1},\mathbf{1})_{-2}$ is also present. (c) The three-loop model from Ref.~\cite{Krauss:2002px}, which has two singlet charged scalars $S_{1,2} \sim (\mathbf{1},\mathbf{1})_{-1}^{}$, and a right-handed Majorana neutrino $N_R \sim (\mathbf{1},\mathbf{1})_{0}$.}
\label{fg:neutrino_mass_model}
\end{figure}

Similar to the doublet and triplet cases, the singlet charged scalar also exists in many NP models. The quantum number of the singlet charged scalar $S$ under the SM gauge group is $(\mathbf{1},\mathbf{1})_{-1}^{}$, where the first and second numbers inside the parenthesis indicate the quantum numbers of $SU(3)_{\mathrm{C}}^{}$ and $SU(2)_{\mathrm{L}}$, respectively, and the subscript denotes the hypecharge. It carries a unit of electric charge under the convention of $Q = I_3 + Y$. Therefore, the singlet charged scalar can be the right-handed slepton in supersymmetric models~\cite{Martin:1997ns}, or it can exist in lots of radiative neutrino mass models at one-loop~\cite{Zee:1980ai}, two-loop~\cite{Aoki:2010ib,Babu:1988ki}, or three-loop ~\cite{Ahriche:2013zwa,Krauss:2002px} level. Some of the radiative neutrino mass models are depicted in Fig.~\ref{fg:neutrino_mass_model}. See Refs.~\cite{Guella:2016dwo,Chekkal:2017eka} for recent works on searching for the singlet charged scalar within neutrino mass models. For simplicity, here we assume that the singlet charged scalar is the only light NP particule with the mass around the electroweak scale, while other possible NP particles are heavy and have little impact on the phenomenology at low energy.

The renormalizable Lagrangian involving $S$ is
\begin{eqnarray} \label{eq:dim_4}
\mathcal{L}_S^{\mathrm{dim-4}} &\supset&  (D_\mu S)^\dagger D^\mu S-m_S^2 |S|^2-\frac{\lambda_S}{2}|S^\dagger S|^2\nn\\
&&-\lambda_{SH} S^\dagger S H^\dagger H + \big( f_{\alpha \beta} \overline{\ell}_{\mathrm{L}\alpha}^{} \ell_{\mathrm{L}\beta}^c S + \mathrm{h.c.} \big) ,
\end{eqnarray}
where $m_S^{}$ is the mass of the singlet charged scalar, and $\lambda_S^{}$ and $\lambda_{SH}^{}$ are the quartic couplings describing the self-interaction of $S$ and the interaction with the SM Higgs doublet $H= (H^+, H^0)^T$, respectively. We denote $\ell_{\mathrm{L}}$ as the lepton doublet in the SM while $\ell_{\mathrm{L}}^c$ its charge conjugated field. The coupling $f_{\alpha\beta}$ (for $\alpha, \beta = 1, 2, 3$) needs to be anti-symmetric on the family indices $\alpha$ and $\beta$ due to Fermi-Dirac statistics. The presence of the interaction $\bar{\ell}_{\mathrm{L}}^{} \ell_{\mathrm{L}}^c S$ would induce large charged lepton flavor violation~\cite{McLaughlin:1999rr}, which is severely constrained by the muon $(g-2)$~\cite{Queiroz:2014zfa,Lindner:2016bgg} and lepton rare decay data~\cite{TheMEG:2016wtm,Olive:2016xmw}. Following Refs.~~\cite{McLaughlin:1999rr,Lavoura:2003xp}, we obtain the constraints $|f_{e\tau} f_{\mu\tau}|, |f_{e\mu} f_{\mu\tau}|,  |f_{e\mu} f_{e\tau}| \lesssim \mathcal{O}(10^{-5})$. For simplicity, we assume that all $f_{\alpha\beta}$'s are highly suppressed such that the $\bar{\ell}_{\mathrm{L}}^{} \ell_{\mathrm{L}}^c S$ term is negligible in comparison with the dimension-5 operators discussed below.

At dimension-5, there exist three types of operators that involves both the scalar $S$ and the SM fields. Some operators are redundant and can be reduced with the help of the equations of motion (EOMs).

The first type involves two singlet charged scalars and two SM fermion fields, i.e., $\bar e_{\rm R}^{} e^c_{\rm R} SS$, with $e_\mathrm{R}^{}$ standing for the right-handed lepton field. The operator is irrelevant for searching for singlet charged scalar at the Large Hadron Collider (LHC)~\footnote{But it can be probed, if an $e^-e^-$ collider were built~\cite{Wang:2016eln}.}.  

The second type of dimension-5 operators is made of two SM fermions and one singlet charged scalar field, i.e., $\overline{Q}_{\rm L}^{} H u_{\rm R}^{}S$, $\overline{Q}_{\rm L}\widetilde{H} d_{\rm R}^{} S^\dagger$ and $\overline{\ell}_{\rm L}^{} \widetilde{H} e_{\rm R}^{} S^\dagger$, where $\widetilde{H} = i\sigma_2^{} H^*$, $Q_{\mathrm{L}}^{}$ is the left-handed quark doublet, and $u_{\mathrm{R}}^{}$ and $d_{\mathrm{R}}^{}$ are respectively the up- and down-type right-handed quark singlets. Although one could write down other fermionic operators with one derivative, they can be reduced to the above three operators using EOMs. 

The last type of dimension-5 operators is purely bosonic. After removing those operators that have two derivatives' acting on a single field (hence reducible with EOMs), we end up with only two options:  
\beq
(D_\mu \widetilde{H})^\dagger (D^\mu H)S^\dagger,\quad\widetilde{H}^\dagger D^\mu HD_\mu S.
\eeq
The first operator vanishes because of Bose statistics. The second operator $\mathcal{O}_B^{} \equiv \widetilde{H}^\dagger D^\mu HD_\mu S$, when expanding out the covariant derivatives, yields
\beq \label{eq:OB_interaction_before_EWSB}
\frac{gg'}{\sqrt{2}}W_\mu^+B^\mu H^0H^0S,\quad -\frac{ig}{\sqrt{2}}W_\mu^+ H^0H^0\partial^\mu S,
\eeq
where $W_\mu^+$ ($B_\mu^{}$) is the gauge field of the SM gauge group $SU(2)_{\mathrm{L}}^{}$ ($U(1)_\mathrm{Y}^{}$), with the gauge coupling dentoed as $g$ ($g^\prime$).  After the spontaneous symmetry breaking of $SU(2)_{\mathrm{L}}^{}$, the Higgs field develops a vacuum expectation value (VEV) as $H^0\to(v+h)/\sqrt{2}$, where $v$ and $h$ represent the electroweak VEV and the physical Higgs field, respectively. The interactions in Eq.~(\ref{eq:OB_interaction_before_EWSB}) then lead to pure bosonic decay modes of $S$, if kinematically allowed, e.g., 
\begin{align}
\label{fake_decay}
& S^- \to W^- Z(\gamma), \quad W^- h, \quad W^- hZ(\gamma),
\end{align}
where $Z$ is the neutral gauge field in the SM electroweak interactions, and $\gamma$ stands for the photon field. Here, we denote $S$ as $S^-$ to keep the electric charge manifest (also $S^\dagger$ as $S^+$ hereafter).

On the other hand, a relation,
\begin{align}
0&=(D_\mu \widetilde{H})^\dagger(D^\mu H)S = D_\mu(\widetilde{H}^\dagger D^\mu H)S -\widetilde{H}^\dagger (D_\mu D^\mu H) S \nn\\
&=\partial_\mu(\widetilde{H}^\dagger D^\mu H S) - \widetilde{H}^\dagger D^\mu H D_\mu S - \widetilde{H}^\dagger  (D_\mu D^\mu H ) S,\nn
\end{align}
shows that $\widetilde{H}^\dagger D^\mu H D_\mu S$ is equivalent to $\widetilde{H}^\dagger  (D_\mu D^\mu H ) S$. The latter operator can be reduced to the second type of operators involving fermions using the EOM of the Higgs field. Apparently, none of those fermion-type operators could yield the decay modes listed in Eq.~(\ref{fake_decay}).  We then confront with a paradox that two equivalent operators yield different decay modes!

The paradox is resolved as follows. After the spontaneous symmetry breaking, the operator $\mathcal{O}_B^{}$ yields the two-point vertices of
\beq
\frac{v}{\sqrt{2}}\left(\partial_\mu H^+\partial^\mu S^- - \frac{igv}{2}W^+_\mu \partial^\mu S^- \right)+\rm{h.c. ~.}
\eeq
The first term indicates a kinetic mixing between the charged Goldstone component $H^+_{}$ of the SM Higgs field and the scalar $S^+$, and the second term suggests revisiting gauge-fixing in the conventional $R_\xi$ quantization procedure.  After proper treatments of field redefinition and gauge fixing, we find that those ``fake" decay channels in Eq.~(\ref{fake_decay}) vanish~\footnote{A similar conclusion of vanishing decay modes from high-dimensional operators is pointed out in Ref.~\cite{Bauer:2016zfj}, and the gauge fixing issue in the SM effective field theory is investigated in Ref.~\cite{Dedes:2017zog} recently.}. To illustrate this, we begin with the relevant terms in the Lagrangian
\begin{eqnarray}
\label{HDHDS*}
\mathcal{L} &\supset& (D_\mu H)^\dagger (D^\mu H) + D_\mu S^+ D^\mu S^- \nonumber \\
& & - m_S^2 S^+S^- +\left( \frac{C_5}{\Lambda}\mathcal{O}_{B}+\text{h.c.}\right),
\end{eqnarray}
where $C_5$ denotes the Wilson coefficient of $\mathcal{O}_B^{}$, and $\Lambda$ stands for the cutoff scale of effective field theory. Expanding out the covariant derivatives associated with the Higgs field and the scalar $S$,  we obtain two-point vertices as follows:
\begin{eqnarray}
\mathcal{L}^{\text{2-point}} & \supset & \partial_\mu H^+ \partial^\mu H^- +\partial_\mu S^+ \partial^\mu S^- - m_S^2 S^+S^- \nonumber \\
& & -\frac{i gv}{2}W_\mu^+ \partial^\mu H^-+\text{h.c.} \nn\\
& &+\frac{C_5}{\Lambda}\frac{v}{\sqrt{2}} \left (\partial_\mu H^+\partial^\mu S^--\frac{igv}{2}W^+_\mu \partial^\mu S^- \right )+\text{h.c.}.\nn
\end{eqnarray}
We then deal with the gauge fixing issue first. Defining $s_\theta=C_5v/(\sqrt{2}\Lambda)$ as a small expansion number that would be kept only at linear order, we rotate the fields of $S^\pm$ and $H^\pm$ as
\beq
\begin{pmatrix}G_1^\pm\\ G_2^\pm\end{pmatrix}=
\begin{pmatrix}
1 & s_\theta \\ -s_\theta & 1
\end{pmatrix}
\begin{pmatrix}
H^\pm \\ S^\pm
\end{pmatrix},
\eeq
which transforms the above two-point interactions into
\begin{eqnarray}\label{2-point}
\mathcal{L}^{\text{2-point}}& \supset & \begin{pmatrix}
\partial_\mu G_1^+ & \partial_\mu G_2^+
\end{pmatrix} \begin{pmatrix}
1 & s_\theta^{} \\
s_\theta^{} & 1
\end{pmatrix}
\begin{pmatrix}
\partial_\mu G_1^- \\  
\partial_\mu G_2^-
\end{pmatrix} \nonumber \\
& & - \begin{pmatrix}
G_1^+ & G_2^+
\end{pmatrix}
\begin{pmatrix}
0 & s_\theta^{} m_S^2 \\
s_\theta^{} m_S^2 & m_S^2
\end{pmatrix}
\begin{pmatrix}
G_1^- \\ G_2^-
\end{pmatrix} \nonumber \\
& & -i\frac{gv}{2}W_\mu^+\partial^\mu G^-_1+\rm{h.c.}~.
\end{eqnarray}
Now $G_1^\pm$ is the only field coupling with the gauge boson $W_\mu^\pm$, and therefore choosing proper gauge fixing terms as 
\begin{align}
\mathcal{L}_{\rm GF}\supset&-\frac{1}{\xi}(\partial^\mu W_\mu^-+i\xi\frac{gv}{2}G_1^-)(\partial^\mu W_\mu^+-i\xi\frac{gv}{2}G_1^+),
\end{align}
with $\xi$ being the gauge fixing parameter, cancels out the term $W_\mu^+\partial^\mu G^-_1$ in Eq.~(\ref{2-point}), so that the issue of gauge fixing is resolved. Next, we diagonalize the kinetic mixing matrix of $G_{1,2}^\pm$, make it canonical and finally diagonalize the mass matrix. The whole procedure results in a field redefinition of
\beq
\begin{pmatrix}G^\pm_1\\ G^\pm_2\end{pmatrix}=
\begin{pmatrix}
1 & 0 \\ -s_\theta & 1
\end{pmatrix}
\begin{pmatrix}
S^\pm_1 \\ S^\pm_2
\end{pmatrix},
\eeq
which leads to
\beq
\mathcal{L}\supset\partial_\mu S_1^+ \partial^\mu S_1^- +\partial_\mu S_2^+ \partial^\mu S_2^- - \xi m_W^2S_1^+S_1^-  - m_S^2 S_2^+ S_2^-~. \nn
\eeq
The physical particle $S_2^\pm$ with mass $m_S^{}$ is then seperated from the Goldstone field $S_1^\pm$ with unphysical gauge-dependent mass $\sqrt{\xi}m_W$. In all, we obtain a transformation matrix (not unitary) that turns the fields in the interaction states to those in physical states,
\beq
\begin{pmatrix}H^\pm\\ S^\pm\end{pmatrix}=
\begin{pmatrix}
1 & -s_\theta \\ 0 & 1
\end{pmatrix}
\begin{pmatrix}
S^\pm_1 \\ S^\pm_2
\end{pmatrix}. 
\eeq
With the above transformation, below we prove the bosonic modes of $S$ decay listed in Eq.~(\ref{fake_decay}) indeed vanish.

First, consider $S_2^\pm\to W^\pm(Z/\gamma)$. This channel can be read out from $\mathcal{O}_B^{}$ via 
\beq
\frac{C_5}{\Lambda}\mathcal{O}_B\supset s_\theta\frac{gg'}{2}vW_\mu^+(c_WA^\mu-s_WZ^\mu)S_2^-+\rm{h.c.}~.
\eeq 
Using $H^\pm=S_1^\pm-s_\theta S_2^\pm$, we also obtain a similar term from the SM sector,
\beq
(D_\mu H)^\dagger D^\mu H\supset -s_\theta\frac{gg'}{2}vW_\mu^+(c_WA^\mu-s_WZ^\mu)S_2^-+\rm{h.c.}.
\eeq
The two terms cancel out each other exactly.

Second, consider $S_2^\pm\to W^\pm h$. The channel originates from the operator
\beq
\frac{C_5}{\Lambda}\mathcal{O}_B\supset -igs_\theta W_\mu^+h\partial^\mu S_2^-+\rm{h.c.}.
\eeq
The counter part from the SM sector is given by
\beq
(D_\mu H)^\dagger D^\mu H\supset \frac{ig}{2}s_\theta W_\mu^+(h\partial^\mu S_2^--\partial^\mu h S_2^-)+\rm{h.c.}~.
\eeq
Their sum reads
\beq\label{S-W-h}
-\frac{ig}{2}s_\theta W_\mu^+\partial^\mu(hS_2^-)+\text{h.c.}=\frac{ig}{2}s_\theta(\partial^\mu W_\mu^+)hS_2^-+\rm{h.c.},
\eeq
which is zero for an on-shell $W^\pm$. Therefore, there is no $S_2^\pm\to W^\pm h$ channel as well. 

Finally, consider $S_2^\pm\to W^\pm h Z(\gamma)$. For convenience we choose the unitary gauge ($\xi\to\infty$), in which only physical fields such as $S_2^\pm$, $W^\pm$ and $h$ exist. Then, the  four-point contact interaction presented in
\beq
\frac{C_5}{\Lambda}\mathcal{O}_B\supset\frac{gg's_\theta}{2}h(S_2^+W_\mu^-+ S_2^-W_\mu^+)(c_W^{} A^\mu -s_W^{} Z^\mu),\nn
\eeq
where $c_W^{} = \cos \theta_{W}^{}$ and $s_W^{} = \sin \theta_{W}^{}$ with $\theta_W^{}$ being the Weinberg angle,  contributes to the decay. That leads to an amplitude of
\beq
i\mathcal{M}_1=\frac{igg's_\theta}{2}\epsilon^*_W\cdot(c_W\epsilon^*_\gamma -s_W\epsilon^*_Z),
\eeq
where  $\epsilon$'s are the polarization vectors of the external fields. Besides, there is a second diagram in which $S_2^\pm$ splits into $h$ and $W^\pm$ through the vertex in Eq.~(\ref{S-W-h}) and $Z/\gamma$ radiates from the $W$ boson. Defining $q=p_W^{}+p_{Z/\gamma}^{}$ with $p_W^{}$ and $p_{Z/\gamma}^{}$ being the extermal momenta of $W$ and $Z/
\gamma$, respectively, the second diagram yields an amplitude of
\begin{align}
i\mathcal{M}_2&=-\frac{igs_\theta}{2}q^\sigma\frac{i(-g_{\sigma\nu}+\frac{q_\sigma q_\nu}{m_W^2})}{q^2-m_W^2} ig(c_W\epsilon^*_{Z\rho}+s_W\epsilon^*_{\gamma\rho})\epsilon^*_{W\mu}\nn\\
\times\big[(q&+p_W)^\rho g^{\mu\nu}-(p_{Z/\gamma}+q)^\mu g^{\rho\nu}-(p_W-p_{Z/\gamma})^\nu g^{\rho\mu}\big]\nn\\
&=-\frac{igg's_\theta}{2}\epsilon^*_W\cdot(c_W\epsilon^*_\gamma -s_W\epsilon^*_Z).
\end{align}
The total amplitude of $S_2^\pm\to W^\pm(Z/\gamma)h$ thus vanishes. Other purely bosonic decay channels can be shown to be absent similarly. Hence, we can indeed trade $\mathcal{O}_B^{}$ for a combination of fermion-type operators, which only lead to fermionic decay modes of $S^\pm$. The Equivalence Theorem in effective field theory~\cite{Arzt:1993gz,Wudka:1994ny} is thus explicitly verified in the effective filed theory with the scalar $S^\pm$.

One may wonder what would happen if the operator $\mathcal{O}_B^{}$ were kept while eliminating one of the fermion-type operators, say, $\overline{\mu}_\mathrm{L}^{} \widetilde{H} \mu_\mathrm{R}^{} S^\dagger$, as according to Equivalence Theorem in effective field theory there exists the freedom of choosing which set of operators are independent. Removing $\overline{\mu}_\mathrm{L}^{} \widetilde{H} e_\mathrm{R}^{} S^\dagger$ seems that we do not have the decay mode of $S^+\to\mu^+\nu_\mu$ at a first glance. However, due to the fact that the $H^+_{}$ component of the SM Higgs field now has some $S_2^+$ contribution, so that from the SM Yukawa interaction 
\beq
 \bar\nu_{\mu\rm L}H^+\mu_{\rm R}^{}= \bar\nu_{\mu\rm L}(S_1^+-s_\theta S_2^+)\mu_{\rm R}^{},\nn
\eeq
the decay channel $S^+\to\mu^+\nu_\mu$ resurrects. 

In short, there are only {\it four} independent dimension-5 operators (for one generation),
\begin{eqnarray}
\bar e_{\rm R}^{} e^c_{\rm R} SS, \quad \overline{Q}_{\rm L}^{} H u_{\rm R}^{}S, \quad \overline{Q}_{\rm L}\widetilde{H} d_{\rm R}^{} S^\dagger, \quad \overline{\ell}_{\rm L}^{} \widetilde{H} e_{\rm R}^{} S^\dagger,\nn
\end{eqnarray}
which give rise to the following major modes of $S$ decay, if kinematically allowed,
\beq
\label{eq:decay_channel}
S^-\rightarrow \begin{dcases*}
    e^-\bar{\nu}, \mu^-\bar{\nu}, \tau^-\bar{\nu},\\
    d\bar{u}, s\bar{c}, b\bar{t}.
\end{dcases*}
\eeq
The three-body decays are not considered as they suffer from huge suppression of phase space. 

\section{The LEP Constraint} 

After specifying the dominant decay modes of $S^\pm$, we are ready to discuss constraints from the LEP data. The constraints from electroweak oblique parameters are weak. According to Ref.~\cite{Grimus:2008nb}, at one-loop level, $S^\pm$ only affects the $S$-parameter slightly, and does not contribute to the $T$-parameter. For example, a deviation of $\delta S=-0.001$ is obtained for $m_S=70$ GeV, while the current experimental value of $S$-parameter is $0.05\pm0.10$~\cite{Patrignani:2016xqp}. The scalar $S^\pm$ is produced in pair through the Drell-Yan channel $e^+e^-(q\bar{q})\to \gamma/Z\to S^+S^-$ at the LEP (LHC), with subsequent decays into pairs of leptons or quarks. No experiment has been carried out on searching for the singlet charged scalar yet, but a few channels of NP searches exhibit the same event topology and thus can be used to constrain $S^\pm$. 

\begin{figure} 
\centering
\includegraphics[scale=0.6]{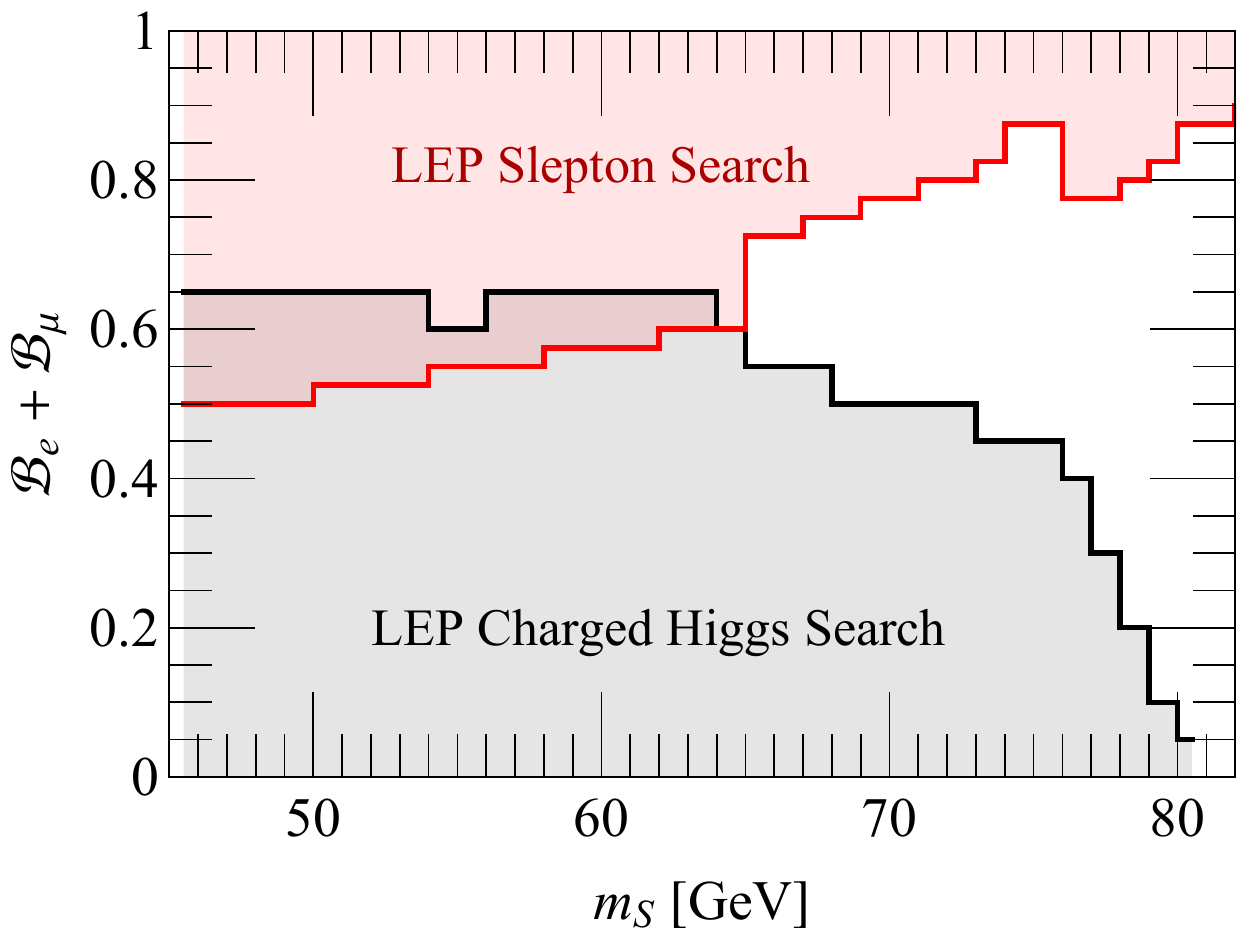}
\caption{Exclusion limits on the singlet charged scalar obtained from the LEP charged Higgs pair search (gray region) and the LEP slepton pair search (red region) at the $95\%$ confidence level. }
\label{fg:combined_limit_LEP}
\end{figure}

The LEP collaborations have searched for the charged Higgs boson pairs in the THDM. The analysis focuses on three types of final states, $\tau^+_{} \nu_\tau^{} \tau^-{} \bar{\nu}_\tau^{}$, $c\bar{s} \tau^-_{} \bar{\nu}_\tau^{}$($\bar{c}s \tau^+_{} \nu_\tau^{}$) and $c\bar{s}\bar{c}s$, under the assumption that $\mathcal{B}(\mathcal{H}^+_{} \rightarrow c \bar{s} ) + \mathcal{B}(\mathcal{H}^+_{} \rightarrow \tau^+_{} \nu_\tau) = 1$, where $\mathcal{B}$ stands for the decay branching ratio and $\mathcal{H}^+_{}$ denotes the charged Higgs field in the THDM. The assumption is valid in the THDM as the couplings of the charged Higgs $\mathcal{H}^\pm$ to the SM fermions are proportional to the SM fermion masses. A combined analysis of all the LEP data shows that the charged Higgs below 80 GeV in the THDM is excluded at the $95\%$ confidence level~\cite{Abbiendi:2013hk}. Such a lower bound is not applicable to the scalar $S^\pm$, as $S^\pm$ could sizably decay into electrons and muons. Nevertheless, for a given value of $m_S^{}$ one can still employ the LEP limits in Ref.~\cite{Abbiendi:2013hk} to obtain a constraint on $(\mathcal{B}_e+\mathcal{B}_\mu)$ , with $\mathcal{B}_\ell$ standing for the decay branching ratio to the charged lepton $\ell$ ($\ell = e, \mu$). In Figure~\ref{fg:combined_limit_LEP} we show such excluded parameter space as the gray shaded region. Hence a singlet charged scalar with mass as low as $45~\mathrm{GeV}$ is still allowed by the LEP charged Higgs search as long as $(\mathcal{B}_e + \mathcal{B}_\mu) \gtrsim 0.7$.

Another constraints on $(\mathcal{B}_e + \mathcal{B}_\mu)$ can be obtained from the slepton search at the LEP, where the sleptons are produced in pair via the $s$-channel Drell-Yan process.  Only the same flavor dilepton final state is considered in the LEP experiments, i.e., $e^+e^-+\met$ and $\mu^+\mu^- + \met$, where $\met$ denotes the missing transverse momentum. The null result is translated into an upper bound on the cross section of the slepton pair production on the condition that the slepton decays via only the electron mode or the muon mode. In the translation we adopt the LEP constraints in the case of massless neutralino, as the neutrino in the $S^\pm$ decay is massless. 
We end up with two constraints as follows: 
\bea
&&\sigma(S^+S^-)\mathcal{B}_e^2 \leq \sigma(\tilde{\ell}^+\tilde{\ell}^-)_{e},\nn\\
&&\sigma(S^+S^-)\mathcal{B}_\mu^2 \leq \sigma(\tilde{\ell}^+\tilde{\ell}^-)_{\mu},
\eea
where $\sigma(S^+S^-)$ denotes the inclusive cross section of the charged scalar pair production while $\sigma(\tilde{\ell}^+\tilde{\ell}^-)_{e,\mu}$ represents the upper limit on the cross sections of the slepton pair production obtained in the electron mode and muon mode at the LEP~\cite{LEP_combine}, respectively. 
We vary $\mathcal{B}_e$ and $\mathcal{B}_\mu$ independently to constrain $(\mathcal{B}_e+\mathcal{B}_\mu)$. In the study we are interested in the allowed parameter space of the scalar $S$, therefore, we obtain the maximal upper limit of $(\mathcal{B}_e+\mathcal{B}_\mu)$ for each $m_S^{}$; see the red shaded region in Figure~\ref{fg:combined_limit_LEP}.  
One then observes that a singlet charged scalar with $m_S^{} \lesssim 65~\mathrm{GeV}$ is excluded by combining both the LEP charged Higgs and slepton search results, while a heavy charged scalar with $m_S^{} > 65~\mathrm{GeV}$ is still allowed (white region). We emphasize that the bound is robust and model-independent.   

\section{The LHC Phenomenology}

\begin{figure}
\centering
\includegraphics[scale=0.6]{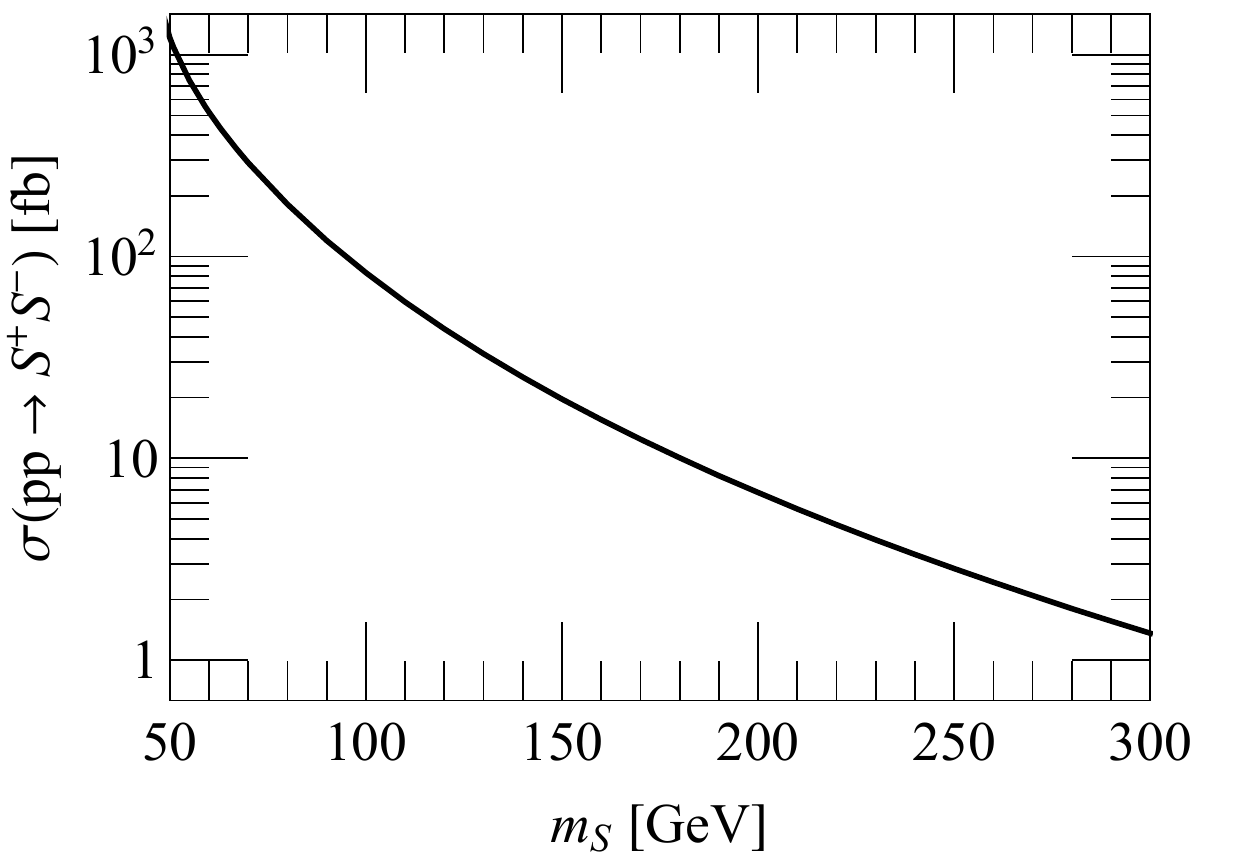}
\caption{Inclusive cross section of the charged scalar pair production as a function of $m_S^{}$ at the 13~TeV LHC. }
\label{fg:xsec}
\end{figure}

At the LHC the scalars $S^\pm$ are produced in pair via the Drell-Yan process $q \bar{q} \rightarrow \gamma/Z \rightarrow S^+ S^-$. Figure~\ref{fg:xsec} displays the inclusive cross section of the charged scalar pair production as a function of $m_S^{}$ at the 13~TeV LHC. The cross section decreases with the $m_S^{}$ dramatically. Although in general $S^\pm$ can decay into both quarks and leptons according to Eq.~(\ref{eq:decay_channel}), because of the large QCD backgrounds at the LHC, searching for $S^\pm$ via the quark decay mode is not promising. Therefore, in this study we focus on the lepton search modes and for simplicity assume the charged scalars does not decay to quarks, i.e., $\mathcal{B}_{\ell}=\mathcal{B}_e+\mathcal{B}_\mu+\mathcal{B}_\tau=1$. Our results, however, can be easily rescaled to incorporate non-zero decay branching ratio to quarks $\mathcal{B}_J^{}$. For instance, the cross section of the dilepton decay mode is rescaled as $\sigma_{\ell J}^{} = \sigma_\ell^{} \left( 1 - \mathcal{B}_J^{} \right)^2$, with $\sigma_{\ell J}^{}$ ($\sigma_\ell^{}$) being the cross section with (without) the dijet decay channels. As the tau lepton exhibits collider signatures different from electrons and muons, we divide the lepton modes into the ``dilepton" mode ($\ell^\pm \ell^{(\prime) \mp} + \met$ with $\ell^{(\prime)} = e, \mu$) and the ``ditau" mode ($\tau^\pm \tau^\mp +\met$), where $\met$ denotes the missing transverse momentum arising from the two invisible neutrinos.

\subsection{The LHC constraint}

First consider the existing charged Higgs search at the LHC. Since the LEP has already excluded a charged Higgs boson below $80~\mathrm{GeV}$ in the THDM, the LHC experiments target on the charged Higgs boson heavier than $80~\mathrm{GeV}$~\cite{Aad:2014kga,Khachatryan:2015qxa,Khachatryan:2015uua,Aad:2013hla,CMS:2016qoa,Aad:2014kga,Khachatryan:2015qxa,ATLAS:2016grc}, resulting in no constraint on the scalar $S^\pm$ in the case of $m_S^{} < 80~\mathrm{GeV}$.  Moreover, in the searches for a heavy charged Higgs boson, the charged Higgs bosons are always assumed to be produced in association with top quarks (motivated by the THDM), which should be compared with the Drell-Yan production of the scalar $S^\pm$. Therefore, the existing searches for the charged Higgs boson at the LHC yield almost no constraints on either the low or high mass case of $S^\pm$.

On the other hand, the slepton searches at the LHC emcompass the leptonic modes of the scalar $S^\pm$. It is not a surprise as the scalar $S^\pm$ carries exactly the same quantum numbers as the right-hand sleptons under the SM gauge group. Unfortunately, the current searches for sleptons in the dilepton mode also target on the supersymmetric partners of electrons or muons with masses larger than $80~\mathrm{GeV}$ at the LHC~\cite{Aad:2014vma,ATLAS-CONF-2016-096,ATLAS-CONF-2017-039}. That again imposes no constraints on $\mathcal{B}_e$ and $\mathcal{B}_\mu$ for $m_S^{} \lesssim 80~\mathrm{GeV}$. For the ditau mode, there does exist searches~\cite{Aad:2014yka,Aaboud:2017nhr} designed for supersymmetric partners of tau leptons, focusing on both low and high mass ranges. However, the constraints on $\mathcal{B}_\tau$ from those searches are rather weak due to the low integrated luminosity accumulated so far. 

\subsection{Searches in the dilepton mode}

In this subsection we investigate the potential of the $13~\mathrm{TeV}$ LHC on searching for the charged scalar pairs in the dilepton mode with integrated luminosities of $100~\mathrm{fb}^{-1}$, $300~\mathrm{fb}^{-1}$ and $3000~\mathrm{fb}^{-1}$. We use the package \texttt{FeynRules}~\cite{Alloul:2013bka} to generate the UFO model file~\cite{Degrande:2011ua} for the singlet charged scalar. For the event generation, we choose \texttt{MadGraph5\_aMC@NLO}~\cite{Alwall:2014hca} to generate the signal and background events at parton level,  and then employ the packages \texttt{Pythia 6}~\cite{Sjostrand:2006za} and \texttt{Delphes}~\cite{deFavereau:2013fsa} for parton shower and detector simulation, respectively.  Each background process is matched to one additional jet. 

We start with the case of a light charged scalar ($m_S^{}\lesssim 80~\rm{GeV}$). The signal events are characterized by two hard charged leptons ($e$ or $\mu$) and large $\met$. The main background processes are the diboson pair production ($W^+W^-$, $W^\pm Z$ and $ZZ$), the Drell-Yan process $q\bar{q} \rightarrow \gamma/Z \rightarrow \ell^\pm \ell^\mp$ ($\ell = e, \mu, \tau$), $t\bar{t}$ pair production and $tW$ associated production. 
In order to mimic the collider environment, we require that both the signal and background events exhibit exactly two opposite-sign charged leptons ($e$ or $\mu$), and satisfy a set of selection cuts~\cite{ATLAS-CONF-2016-096}:
\begin{align}
&p_T^{\ell_1} > 25~\mathrm{GeV}, &&p_T^{\ell_2} > 20~\mathrm{GeV},  \nn\\
&|\eta^\ell | < 2.5, &&\met>35~{\rm GeV},
\end{align}
where $p_T^{\ell_1}$ ($p_T^{\ell_2}$) denotes the transverse momentum of the leading (sub-leading) charged lepton, and $\eta^\ell$ is the lepton pseudo-rapidity. 
We further demand no hard jet activity in the central region of detector, i.e. vetoing additional jets satisfying $p_T^{j} > 20~\mathrm{GeV}$ and $|\eta^j | < 2.8$~\cite{ATLAS-CONF-2016-096}. 
If the two charged leptons are of the same flavor, we require the invariant mass of dilepton pair ($m_{\ell\ell}$) to be away from the $Z$-pole, i.e. $m_{\ell\ell} \notin [80, 100]~\mathrm{GeV}$. Finally, the hard cut on $\met$ mainly serves to suppress the Drell-Yan background which does not have large missing transverse momentum. For illustration, we choose a benchmark scenario of the signal process, 
\[m_S^{} = 70~\mathrm{GeV},~\mathcal{B}_e = \mathcal{B}_\mu = 0.25~,\] 
and its production rate at the 13~TeV LHC with an integrated luminosity of $100~{\rm fb}^{-1}$ is shown in the second column of Table~\ref{tb:dilepton}. 

We generate the background processes at leading order and rescale them with proper $K$-factors to include higher QCD corrections in accordance to the state-of-the-art theoretical calculations. Specifically, we introduce a factor of $1.2$ in the signal process~\cite{Fuks:2013lya}, $1.5$ in the $WW$ process~\cite{Gehrmann:2014fva,Grazzini:2016ctr,Caola:2015rqy,Bagnaschi:2011tu}, $1.4$ in the $ZZ$ process~\cite{Campbell:1999ah,Aad:2015zqe},  $1.4$ in the $WZ$ process~\cite{Grazzini:2016swo}, $1.2$ in the Drell-Yan process~\cite{Boughezal:2016wmq}, $1.8$ in the $t\bar{t}$ production~\cite{Czakon:2013goa,Ahrens:2011px} and $0.9$ in the $tW$ associated production~\cite{Zhu:2001hw,Campbell:2005bb,Cao:2008af, Frixione:2008yi,Kidonakis:2015nna}, respectively. The numbers of the signal and background events after the selection cuts are shown in the third column of Table~\ref{tb:dilepton}, assuming an integrated luminosity of $100~\mathrm{fb}^{-1}$. It yields a $3.9\sigma$ significance of discovery potential. Since we find that the cut efficiencies of electrons and muons are quite close at the LHC,  the results for other branching ratios of $\mathcal{B}_{e,\mu}^{}$ can be obtained by rescaling. For instance, the required luminosity at the 13 TeV LHC to achieve $5\sigma$ discovery when $m_S^{} = 70~\mathrm{GeV}$ is given by the equation
\begin{eqnarray}
\sqrt{\frac{\mathcal{L}_{\mathrm{disc}}}{100~\text{fb}^{-1}}}\times\left(\frac{\mathcal{B}_e+\mathcal{B}_\mu}{0.25+0.25}\right)^2\times 3.9=5,
\end{eqnarray}
which yields
\begin{eqnarray}
\mathcal{L}_{\mathrm{disc}}^{} =\frac{ 10.27~\mathrm{fb}^{-1}}{ \left( \mathcal{B}_e^{} + \mathcal{B}_\mu^{} \right)^{4}}.
\end{eqnarray}

\begin{table}
\centering
 \caption{ Numbers of the signal ($m_S^{}=70~{\rm GeV}$ with $\mathcal{B}_e=\mathcal{B}_\mu=0.25$) and background events in the dilepton channel at the $13~\mathrm{TeV}$ LHC with an integrated luminosity of $100~{\rm fb}^{-1}$. } \label{tb:dilepton}
\begin{tabular}{c | c | c |c }
\hline
 & No Cut & Selection Cuts & $S/\sqrt{B}$\\
\hline
\tabincell{c}{ $m_S^{} = 70~\mathrm{GeV}$ \\ $\mathcal{B}_e = \mathcal{B}_\mu= 0.25$} & $8.473 \times10^3$ & 1136 & 3.9\\
\hline
 $WW$ & $1.284\times 10^{7}$ & 72700& - \\
 $ZZ$ & $1.560\times 10^{6}$ & 170 & - \\
$WZ$ &  $5.111\times 10^{6}$ & 1210 & -\\
Drell-Yan & $8.264\times 10^{8}$& 6500 &- \\
$t\bar{t}$ & $8.320\times 10^{7}$ & 3540 &- \\
$tW$ & $7.170\times 10^{6}$ &	 2850& - \\
 \hline
 \end{tabular}
\end{table}

\begin{figure*}
\centering
\includegraphics[scale=0.6]{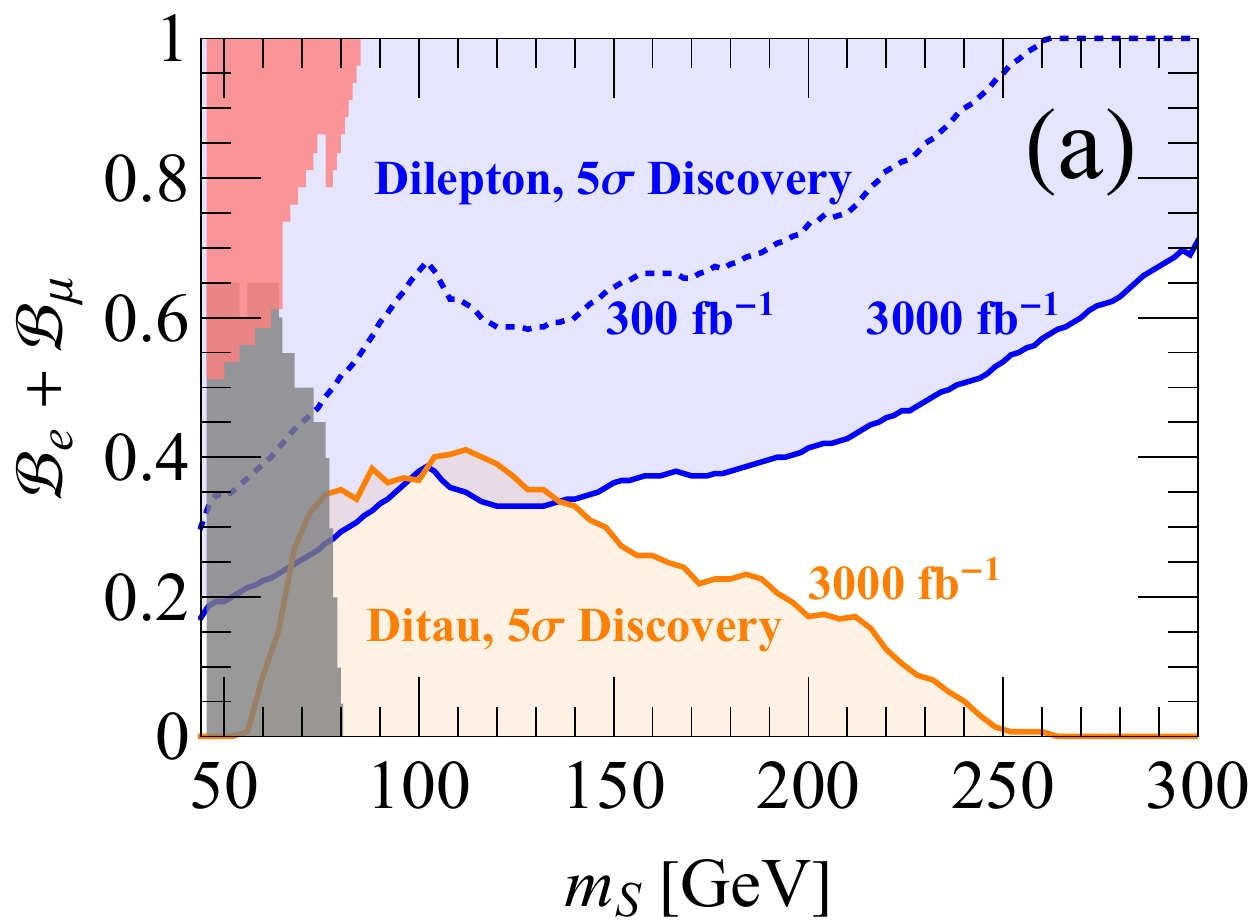}\quad
\includegraphics[scale=0.6]{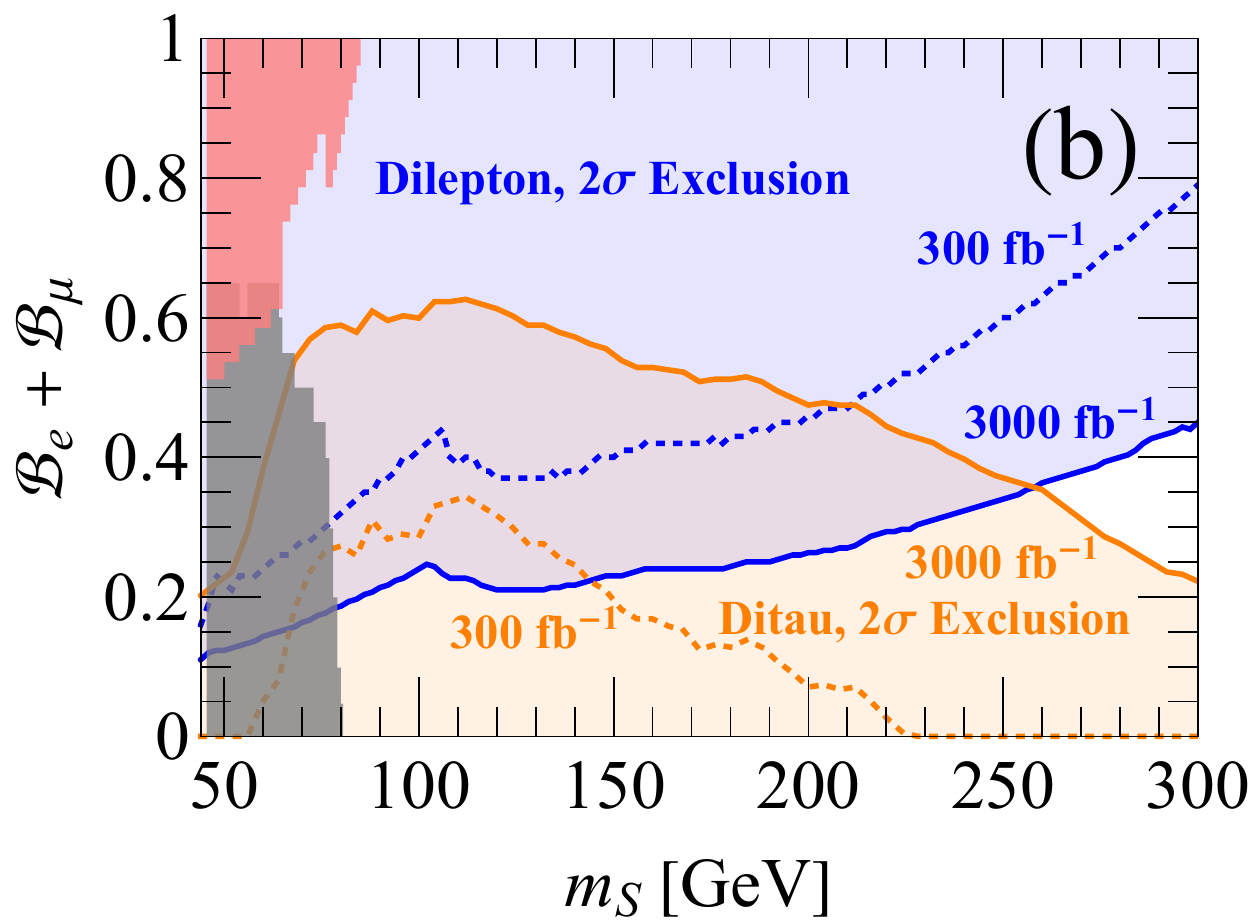}
\caption{Detection prospects for the singlet charged scalar at the 13 TeV LHC via the dilepton and ditau modes: (a) discovery potential; (b) exclusion limits. The gray and red shaded regions are excluded by the charged Higgs and slepton searches at the LEP, respectively.} \label{fg:dilepton_disc_excl}
\end{figure*}

In the case of $m_S^{} > 80~\mathrm{GeV}$ one can further reduce the background from the $WW$ process by employing the $M_{T2}$ variable, which is a function of the momenta of two visible particles and the missing transverse momentum in an event~\cite{Lester:1999tx}. Specifically, $M_{T2}(a,b, \vec{p}_{T}^{\rm~invis})$ is the minimum of the function
\begin{equation}
\max \left\{ M_T(\vec{p}_T^{~a},\not{\!\vec{p}_1}),M_T(\vec{p}_T^{~b},\not{\! \vec{p}_2})\right\} ,
\end{equation}
subject to $\not{\! \vec{p}}_{1}+  \not{\! \vec{p}}_{2}= \vec{p}_{T}^{\rm~invis}$.  Here $a$ and $b$ are the two individual (or clustered) visible states from the parents decay, $\not{\! \vec{p}}_1$ and  $\not{\! \vec{p}}_2$ are the associated missing momenta, and $\vec{p}_{T}^{\rm~invis}$ is the total missing transverse momentum of the event.
The transverse mass $M_{T}$ is defined as 
\begin{equation}
M_{T}(X, \vec{p}_T^{\rm~invis}) = \sqrt{m_X^2 + 2 ( E_T^X \met - \vec{p}_T^X \cdot \vec{p}_{T}^{\rm~invis}) },\nn
\end{equation}
where $X$ denotes the visible particle or cluster,  $E_T^X$ and $\vec{p}_T^X$ represent its transverse energy and transverse momentum, respectively. The mass of $X$ is taken to be $m_X=0$ in this study. The $M_{T2}$ variable has a characteristic feature that its distribution exhibits an endpoint around the true mass of the parent decaying particles, e.g., $m_S^{}$ for the signal process and $m_W$ for the $WW$ background. As a result, demanding $M_{T2} > m_W^{}$ reduces most of the $WW$ background events. 
Recently, the ATLAS collaboration searched for a pair of charginos using exactly the same event topology as our signal process~\cite{ATLAS-CONF-2016-096}.  In our analysis, we follow Ref.~\cite{ATLAS-CONF-2016-096} by considering three $M_{T2}$ cuts, i.e., 
\beq
M_{T2}(\ell_1,\ell_2, \vec{p}_{T}^{\rm~invis}) > 90~\mathrm{GeV},~120~\mathrm{GeV},~150~\mathrm{GeV}.\label{eq:mt2_1}
\eeq
Those hard $M_{T2}$ cuts strongly suppress the SM backgrounds. To minimize the statistical uncertainties in the background, we adopt those simulation results of SM backgrounds from the ATLAS study~\cite{ATLAS-CONF-2016-096} and scale the number of background events linearly with the integrated luminosity. Table~\ref{tb:dilepton_2} shows the numbers of the signal and background events for the case of 
\[m_S^{} = 180~\mathrm{GeV},~\mathcal{B}_e = \mathcal{B}_\mu = 0.5~,\]
in the dilepton channel at the LHC with an integrated luminosity of $100~{\rm fb}^{-1}$. The last column displays the discovery potential of the LHC.

\begin{table}
\centering
\caption{Numbers of the signal ($m_S^{}=180~{\rm GeV}$ with $\mathcal{B}_e=\mathcal{B}_\mu=0.5$) and background events in the dilepton channel at the $13~\mathrm{TeV}$ LHC with an integrated luminosity of $100~{\rm fb}^{-1}$.
The numbers of the background events are adapted from Ref.~\cite{ATLAS-CONF-2016-096}. } \label{tb:dilepton_2}
\begin{tabular}{l | c | c |c}
\hline
 & \tabincell{c}{ $m_S^{} = 180~\mathrm{GeV}$ \\ $\mathcal{B}_e = \mathcal{B}_\mu= 0.5$} & Background & $S/\sqrt{B}$\\
\hline
No Cut & 1208 & -  & -\\
Selection Cuts &  332 & - & - \\
$M_{T2} > 90~\mathrm{GeV}$ & 162 & 959 & 5.2 \\
$M_{T2} > 120~\mathrm{GeV}$ & 95 & 220 & 6.4\\
$M_{T2} > 150~\mathrm{GeV}$ & 35 & 100 & 3.5 \\
\hline
\end{tabular}
\end{table}

Armed with the cut efficiencies of dilepton mode, we are ready to explore the requirements on $\mathcal{B}_e$ and $ \mathcal{B}_\mu$ to reach a $5\sigma$ discovery or a $2\sigma$ exclusion at the 13 TeV LHC. The criteria of $5\sigma$ discovery and $2\sigma$ exclusion are taken to be $S/\sqrt{B} > 5$ and $S/\sqrt{B} > 2$, respectively, where $S$ denotes the number of signal events while $B$ being that of background events. Since the cut efficiencies of electrons and muons are quite close at the LHC, we do not distinguish the charged-lepton flavors in the final state. The analysis on the signal process is not very sensitive to the relative ratio between $\mathcal{B}_e$ and $ \mathcal{B}_\mu$. We thus project the combined results in terms of lower limits on $(\mathcal{B}_e+\mathcal{B}_\mu)$. The mass of scalar $S$ is chosen to be in the range of $m_S^{} \in [44,300]~\mathrm{GeV}$. For each $m_S^{}$ we consider four scenarios, i.e., without $M_{T2}$ cut and applying the three $M_{T2}$ cuts in Eq.~(\ref{eq:mt2_1}) once at a time, and pick up the best sensitivity after combining all the four scenarios. Figure~\ref{fg:dilepton_disc_excl}(a) shows the lower limits of $(\mathcal{B}_e + \mathcal{B}_\mu)$ to reach a $5\sigma$ discovery as a function of $m_S^{}$ at the LHC with an integrated luminosity of $3000~{\rm fb}^{-1}$ (blue) and of $300~{\rm fb}^{-1}$ (blue dashed). The gray and red shaded regions are excluded by the charged Higgs and slepton searches at the LEP, respectively. The corresponding results of the $2\sigma$ exclusion are displayed in Figure~\ref{fg:dilepton_disc_excl}(b). We conclude that, by using the dilepton mode only,  one is able to discover or exclude the scalar $S^\pm$ up to around $80~\mathrm{GeV}$ at the high luminosity LHC. That fills the blind spots in the previous LEP searches.

\subsection{Searches in the ditau channel}

We now turn to the ditau mode. As discussed above, currently there exist several searches for stau scalars $\tilde{\tau}$ at the LHC, which can be used to constrain $\mathcal{B}_\tau$ for both light and heavy $S^\pm$ scalars. Recently, the CMS collaboration searched for a pair of $\tilde{\tau}\tilde{\tau}$ in the $\tau\tau\tilde{\chi}^0\tilde{\chi}^0$ mode ($\tilde{\chi}^0$ being the neutralino), in which the hadronic decay modes of tau leptons are considered~\cite{CMS-PAS-SUS-17-003}. We follow the analysis procedure presented in Ref.~\cite{CMS-PAS-SUS-17-003}, and that enables us to make a quick comparison of our signals to the SM backgrounds given there.

\begin{figure}[b]
\centering
\includegraphics[scale=0.288]{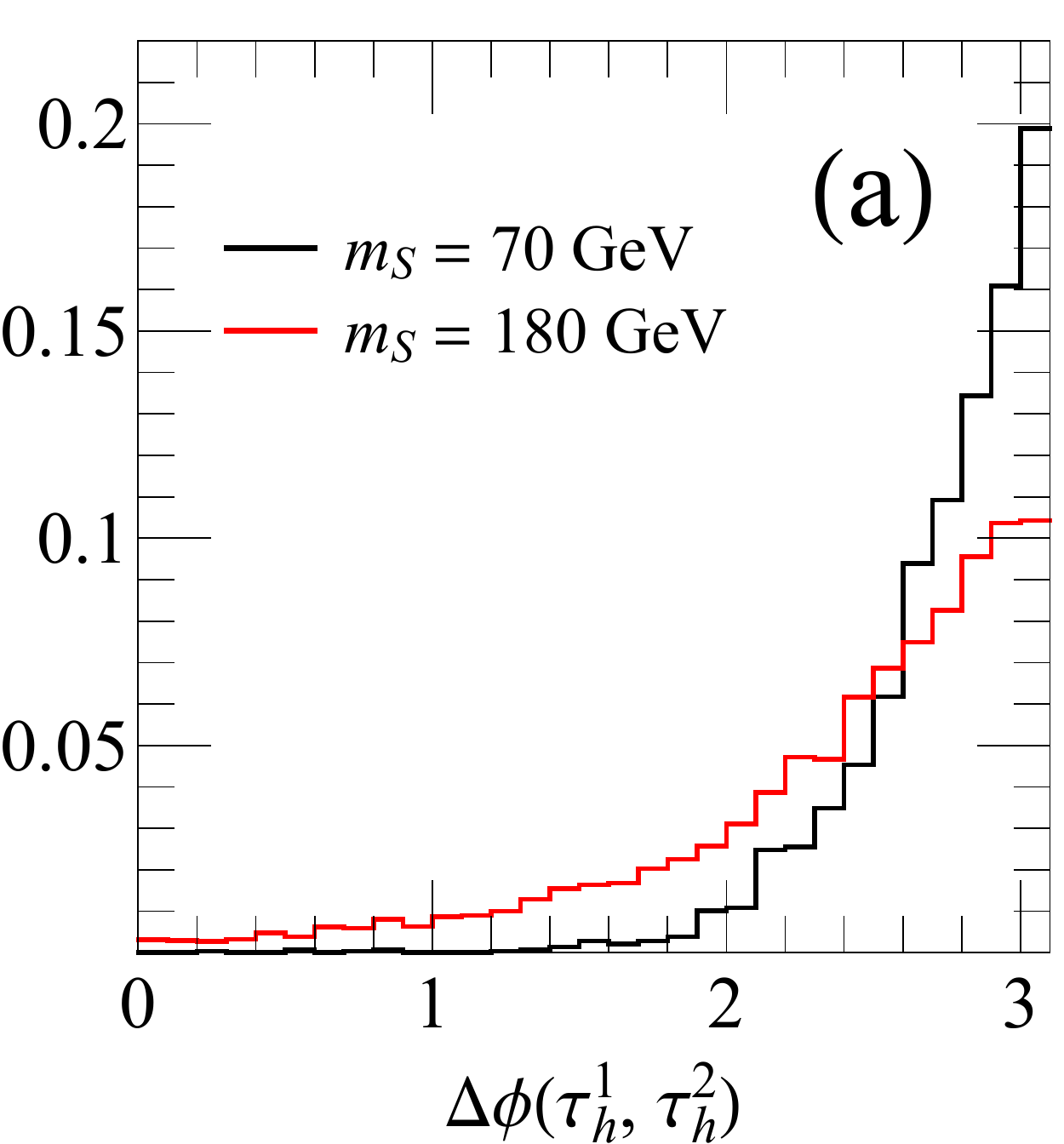} \quad
\includegraphics[scale=0.3]{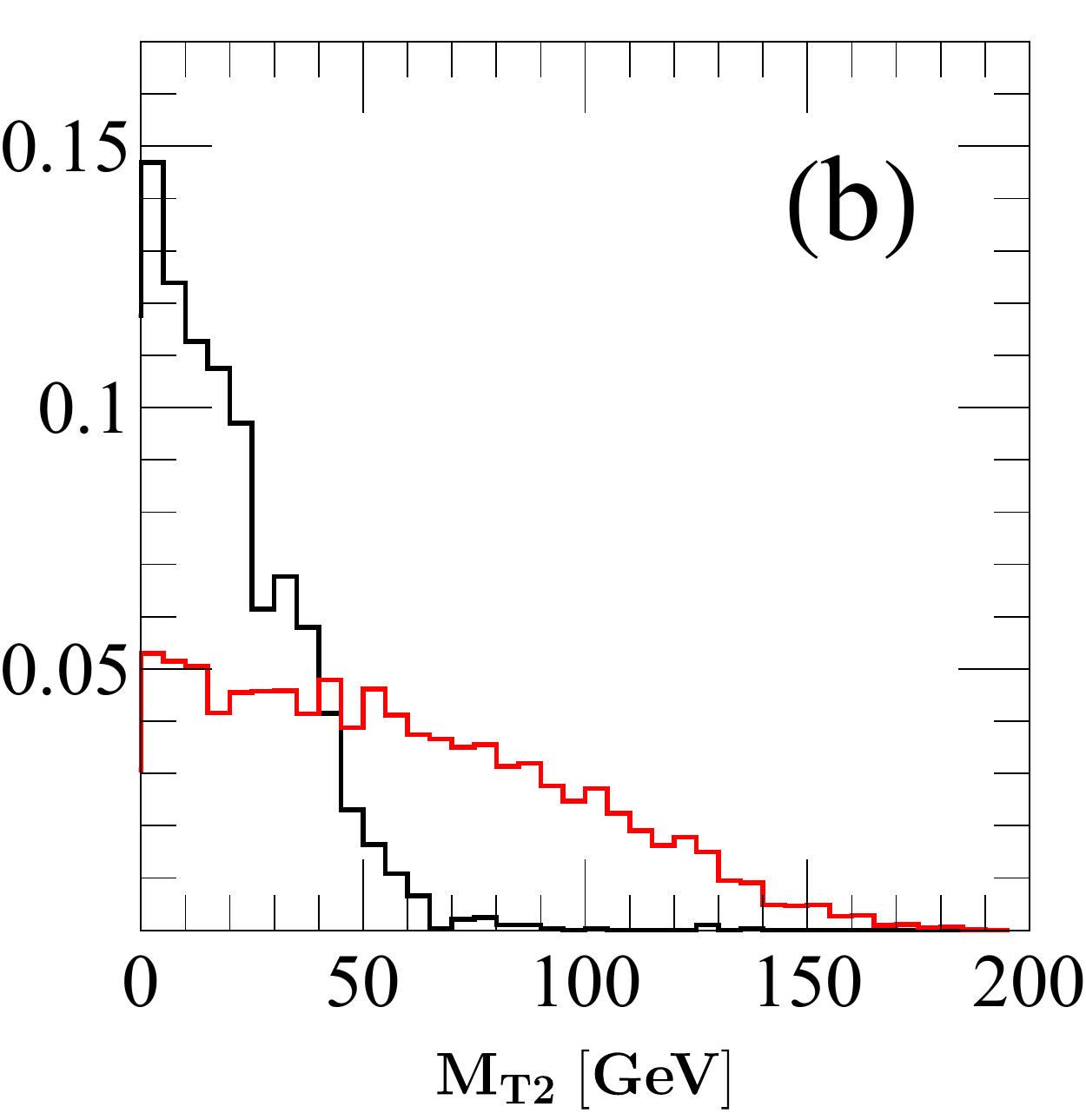} \\
\includegraphics[scale=0.3]{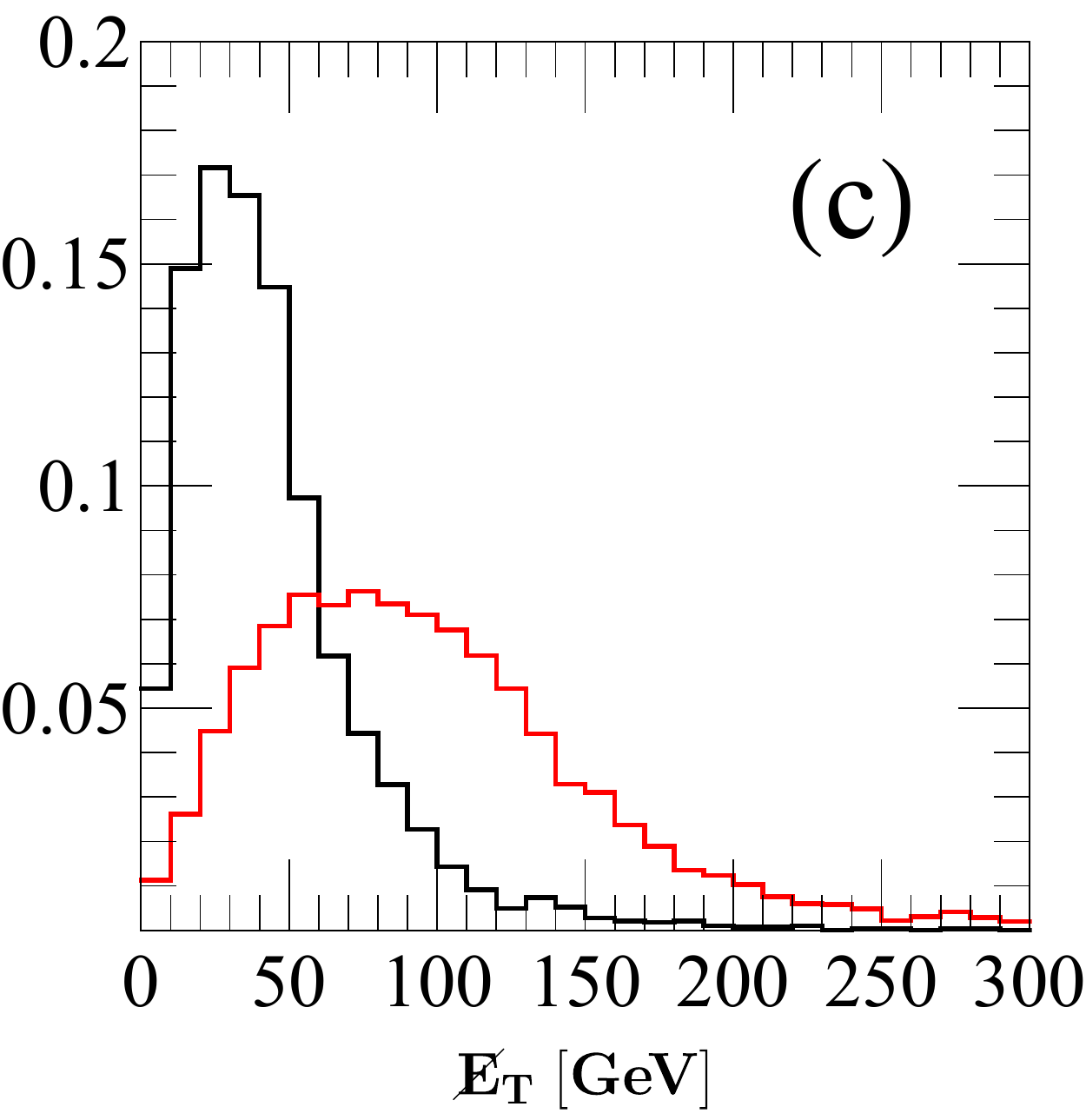} \quad
\includegraphics[scale=0.3]{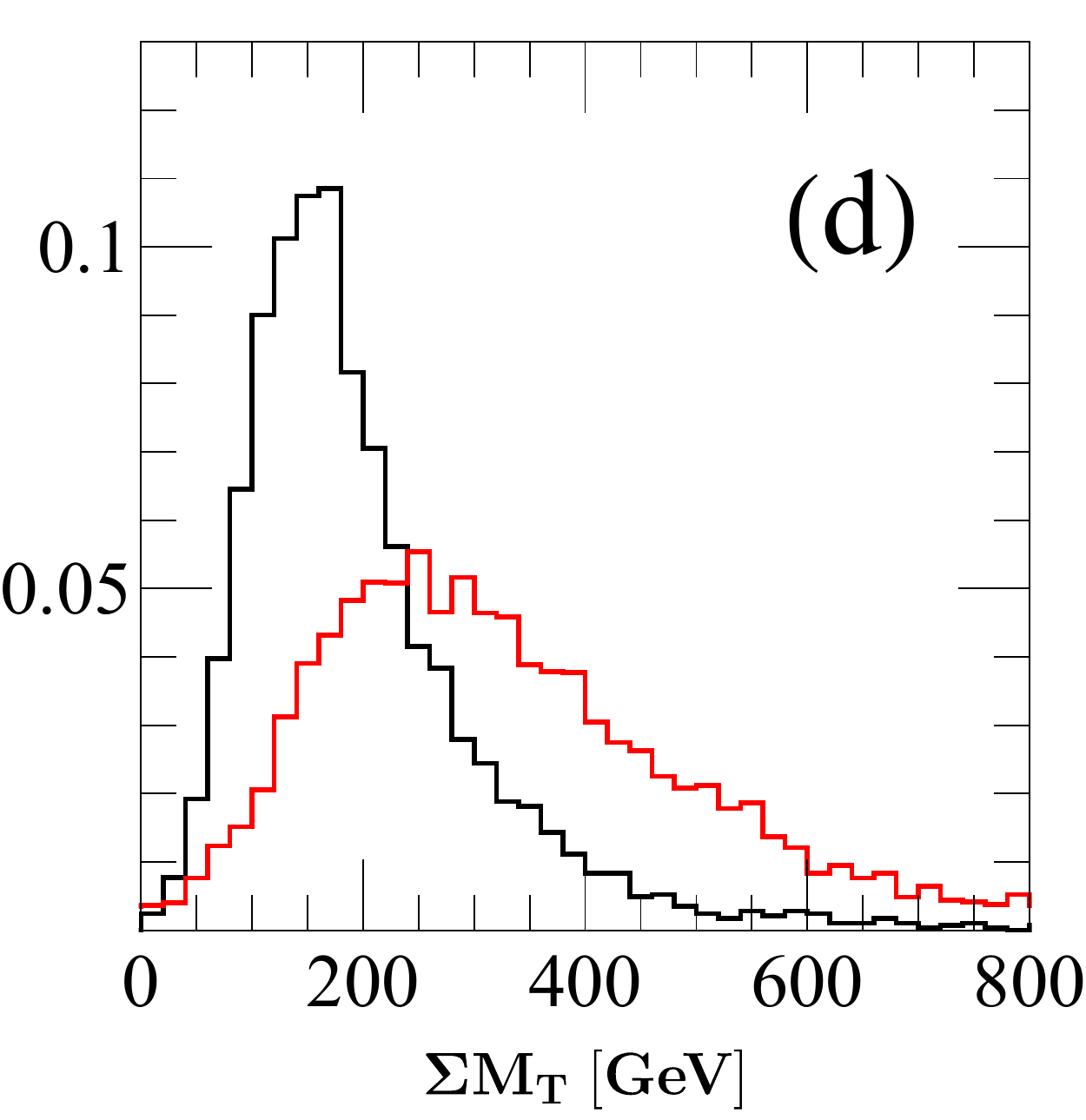}
\caption{Kinematic distributions of (a) $\Delta\phi(\tau_h^1,\tau_h^2)$, (b) $M_{T2}(\tau_h^1,\tau_h^2, \vec{p}_T^{\rm~invis})$, (c) $\met$ and (d) $\Sigma M_T^{}$ for the signal process in the ditau mode.} \label{fg:ditau_dist}
\end{figure}

We first demand exactly two tau-jets satisfying 
\begin{eqnarray}
p_{T}^{\tau_h} > 40~\mathrm{GeV},\quad |\eta^{\tau_h}| < 2.1,
\end{eqnarray}
where $p_{T}^{\tau_h}$ and $\eta^{\tau_h}$ are the transverse momentum and peseudo-rapidity of the tagged tau-jet $\tau_h^{}$, respectively. We further require no hard electrons, muons or jets in the central region; we veto electrons or muons satisfying  
\begin{eqnarray}
p_{T}^\ell > 20~\mathrm{GeV}, \quad |\eta^\ell | < 2.4,
\end{eqnarray}
and QCD jets with 
\begin{eqnarray}
p_{T}^j > 30~\mathrm{GeV}, \quad |\eta^j | < 2.4.
\end{eqnarray}
In the subsequent analysis, three exclusive signal regions (SRs) are considered~\cite{CMS-PAS-SUS-17-003}: 
\begin{itemize}
\item[SR1:] $M_{T2}(\tau_h^1,\tau_h^2, \vec{p}_T^{\rm~invis}) > 90~\mathrm{GeV}$,\\[2mm] 
$|\Delta \phi (\tau_h^1, \tau_h^2) | > 1.5 $~; 
\item[SR2:] $M_{T2}(\tau_h^1,\tau_h^2, \vec{p}_T^{\rm~invis})  \in [40, ~90]~\mathrm{GeV}$,\\[1mm]
$|\Delta \phi (\tau_h^1, \tau_h^2) | > 1.5 $~,\\[2mm]
$\Sigma M_T > 350~\mathrm{GeV}$, \\[2mm]
$\met > 50~\mathrm{GeV}$~;
\item[SR3:]  $M_{T2}(\tau_h^1,\tau_h^2, \vec{p}_T^{\rm~invis})  \in [40, ~90]~\mathrm{GeV}$, \\[2mm]
$|\Delta \phi (\tau_h^1, \tau_h^2) | > 1.5 $~,\\[2mm]
$\Sigma M_T \in [300, 350]~\mathrm{GeV}$,\\[2mm]
$\met > 50~\mathrm{GeV}$~.
\end{itemize}
Here, $\Sigma M_T$ denotes the sum of the transverse masses $M_T$'s of two final state tau-jets,
\beq
\Sigma M_T = M_T(\tau_h^1,  \vec{p}_T^{\rm~invis})+M_T(\tau_h^2, \vec{p}_T^{\rm~invis}),
\eeq
and $\Delta \phi(\tau_h^1, \tau_h^2)$ represents the azimuth angle difference between the two $\tau$-jets. The SR1 aims at a heavy scalar while the SR2 and SR3 are optimized for a light scalar. Figure~\ref{fg:ditau_dist} shows the normalized distributions of $\Delta \phi (\tau_h^1, \tau_h^2)$ (a) and $M_{T2}(\tau_h^1,\tau_h^2, \vec{p}_T^{\rm~invis})$ (b) of the signal process for $m_S^{}=70~{\rm GeV}$ (black) and $m_S^{}=180~{\rm GeV}$ (red) with $\mathcal{B}_\tau = 1$. Owing to the spin correlation effects, the two charged scalars produced through the Drell-Yan channel exhibit a $p$-wave angular distribution such that the matrix element is proportional to $\sin\theta_S^{}$ where $\theta_S^{}$ denotes the polar angle of the charged scalar in the center of mass frame.  Therefore, the two tau leptons from the charged scalar decay are boosted and tend to move back-to-back~\cite{Cao:2003tr}, 
yielding a peak in the $\Delta\phi(\tau_h^1,\tau_h^2)$ distribution around 3 for both light and heavy scalars. Therefore, all the three SRs demand a large $\Delta \phi(\tau_h^1,\tau_h^2)$ to reduce the SM backgrounds. Also, a heavy scalar exhibits a harder $M_{T2}$ distribution than a light scalar, therefore,  the $M_{T2}$ cut used in the SR1 targets at searching for a heavy scalar, while the cuts in the SR2 and SR3 are more effective in searching for a light scalar. Figures~\ref{fg:ditau_dist}(c) and (d) display the normalized distributions of $\met$ and $\Sigma M_T$, respectively. Note that the $\Sigma M_T$ cuts in the SR2 and SR3 are optimized for the $\tilde{\tau}\tilde{\tau}$ searches and may not be the best choices for searching for a pair of singlet charged scalars. Nevertheless, we faithfully follow the CMS collaboration. Table~\ref{tb:ditau} shows the numbers of the signal and background events in the three SRs with an integrated luminosity of $100~\mathrm{fb}^{-1}$. The numbers of signal events shown in the second row (no cut) do not include the branching ratio of tau hadronic decays, while a trigger efficiency of tau jets (60\%) is included. The numbers of all background events in the three SRs, adapted from Ref.~\cite{CMS-PAS-SUS-17-003},  are listed in the fourth column.

\begin{table}
\centering
\caption{Numbers of the signal and background events in the ditau mode at the 13 TeV LHC with an integrated luminosity of $100~\mathrm{fb}^{-1}$. The listed signal regions (SR1, SR2 and SR3) and the corresponding background events in these signal regions are adapted from Ref.~\cite{CMS-PAS-SUS-17-003}. Note that a trigger efficiency of 60\% is included.} \label{tb:ditau}
\begin{tabular}{c | c | c | c}
\hline
\multirow{2}{*}{} & $m_S^{} = 70~\mathrm{GeV}$ & $m_S^{} = 180~\mathrm{GeV}$  & \multirow{2}{*}{Background} \\
& $\mathcal{B}_\tau = 1$ & $\mathcal{B}_\tau = 1$ & \\
\hline
No Cut & 20979 & 725 & - \\
\hline
SR1 & 0 & 4 & 6 \\
SR2 &  5 & 5 & 12 \\
SR3 & 3 & 1 & 10 \\
\hline
\end{tabular}
\end{table}

With the efficiencies of the signal process in the three SRs, we are able to discuss the  search prospects for the scalar $S^\pm$ via the ditau mode. To make a better contact with the search via the dilepton mode, we here assume the decay branching ratio of $S^\pm$ to the quark final state is zero, i.e., $\mathcal{B}_J^{} = 0$. One thus can discuss the discovery or exclusion ability of the ditau mode in terms of an upper limit on $(\mathcal{B}_e + \mathcal{B}_\mu)$, i.e.,  $\mathcal{B}_\tau = 1- (\mathcal{B}_e + \mathcal{B}_\mu)$. Given the presence of three exclusive SRs, we adopt the same combining procedure as in the dilepton mode, except that for each value of $m_S^{}$ we now look for the maximally allowed value of $(\mathcal{B}_e + \mathcal{B}_\mu)$. Figure~\ref{fg:dilepton_disc_excl}(a) shows the discovery region of the ditau mode as a function of $m_S^{}$; the orange shaded region corresponds to the integrated luminosity of $3000~{\rm fb}^{-1}$, while for a luminosity of $300~{\rm fb}^{-1}$ there exist almost no constraints on $(\mathcal{B}_e + \mathcal{B}_\mu)$ (thus not shown).  Assuming that the scalar $S^\pm$ only decays leptonically, we then conclude that at the 13 TeV LHC with the luminosity of $3000~\mathrm{fb}^{-1}$ one can discover the scalar $S^\pm$ up to around $140~\mathrm{GeV}$. 

If null results are reported in the ditau mode, one then can impose constraints on $\mathcal{B}_{\tau}$. Figure~\ref{fg:dilepton_disc_excl}(b) displays the $2\sigma$ exclusion region of the ditau mode at the 13~TeV LHC with the integrated luminosity of $300~{\rm fb}^{-1}$ (orange dashed) and $3000~{\rm fb}^{-1}$ (orange shaded). At the high luminosity LHC, one is able to rule out a singlet charged scalar with mass below 260~GeV after combining the dilepton and ditau modes. That significantly improves the current exclusion limit of 65~GeV from the LEP searches. The conclusion would be weakened if existing a substantial decaying ratio of the scalar $S^\pm$ to quark final states. To incorporate a non-zero $\mathcal{B}_J^{}$, one could shift the upper limits on $(\mathcal{B}_e^{} + \mathcal{B}_\mu^{})$ from the ditau mode down by the amount of $\mathcal{B}_J^{}$, while keeping the same lower limits obtained from the dilepton mode. We note that the actual value of $\mathcal{B}_J^{}$ is very hard to determine at the LHC, however, it becomes possible at the future electron-positron colliders~\cite{Cao:2017ttt}, which would also probe more parameter space, especially those not covered in Figure~\ref{fg:dilepton_disc_excl}. 

\section{Conclusion}

The collider phenomenology of a weak singlet charged scalar ($S^\pm$) is explored in this work. The renormalizable interactions of the singlet scalar $S^\pm$ with the SM leptons are severely constrained by current charged lepton rare decay data. We thus consider the dimension-5 operators built from the singlet charged scalar and the SM fields. Other than the fermionic type of dimension-5 operators, we also identify a bosonic operator $\widetilde{H}^\dagger D^\mu HD_\mu S$, which seemingly leads to the decay modes like $S^\pm \rightarrow W^\pm \gamma$. After performing field redefinitions and introducing gauge fixing terms, we prove that those bosonic decay channels do not exist. 
As a result, the singlet charged scalar can decay only into a pair of leptons or quarks. 

We demonstrate that the existing LEP and LHC constraints do not exclude a singlet charged scalar as light as 65~GeV. Assuming that the singlet charged scalar entirely decays into a pair of leptons, we perform a collider simulation of the scalar pair production in both the $\ell^\pm \ell^{(\prime) \mp}+\met$ mode ($\ell^{(\prime)} =e,\mu$) and $\tau^+\tau^-+\met$ mode. Our study shows that it is very promising to discover the charged scalar up to about $140~\mathrm{GeV}$ at the 13~TeV LHC with an integrated luminosity of $3000~{\rm fb}^{-1}$. If no excess were observed eventually, this same integrated luminosity allows us to exclude a singlet charged scalar with mass below 260~GeV, assuming that only leptonic decay channels are present.

\begin{acknowledgments}
We thank Ran Ding, Yandong Liu, Lian-Tao Wang and Bin Yan for useful discussions. KPX thanks Kai Wang and Jin-Huan He for useful discussions. The work is supported in part by the National Science Foundation of China under Grant Nos. 11275009, 11675002, 11635001 and 11725520, by MOST of ROC under Grant No. MOST106-2112-M-002-003-MY3, and by the China Postdoctoral Science Foundation under Grant No. 2017M610008.
\end{acknowledgments}

\bibliographystyle{apsrev}
\bibliography{reference}
\end{document}